\documentclass[printer]{aa}
\usepackage[varg]{txfonts}
\usepackage{graphicx}
\usepackage{color}
\usepackage{amsbsy,amsmath}
\usepackage{amstext}
\usepackage{natbib}
\usepackage{endnotes}
\newcommand{\be}{\begin{equation}}
\newcommand{\ee}{\end{equation}}
\newcommand{\bea}{\begin{eqnarray}}
\renewcommand{\thefootnote}{\fnsymbol{footnote}}
\newcommand{\eea}{\end{eqnarray}}
\DeclareMathAlphabet\mathbfcal{OMS}{cmsy}{b}{n}

\begin{document}
\title{A self-consistent weak friction model for the tidal evolution of circumbinary planets}
\author{F.A. Zoppetti\inst{1,2}, C. Beaug\'e\inst{1,3}, A.M. Leiva\inst{1} and H. 
Folonier\inst{4}}
\authorrunning{F.A. Zoppetti et al.}
\institute{Universidad Nacional de C\'ordoba. Observatorio Astron\'omico de C\'ordoba, Laprida 
854, C\'ordoba X5000GBR, Argentina.\\
\email{fzoppetti@oac.unc.edu.ar}
\and
Consejo Nacional de Investigaciones Cient\'ificas y T\'ecnicas (CONICET), Argentina.\\
\and
CONICET. Instituto de Astronom\'{\i}a Te\'orica y Experimental, Laprida 854, C\'ordoba 
X5000GBR, Argentina.\\
\and
Instituto de Astronomia Geof\'isica e Ci\^encias Atmosf\'ericas, Universidade de S\~ao 
Paulo, SP 05508-090, Brazil.\\
}

\abstract{
We present a self-consistent model for the tidal evolution of circumbinary planets that is 
easily extensible to any other three-body problem. Based on the weak-friction model, we derive 
expressions of the resulting forces and torques considering complete tidal interactions between 
all the bodies of the system. Although the tidal deformation suffered by each extended mass must 
take into account the combined gravitational effects of the other two bodies, the only tidal 
forces that have a net effect on the dynamic are those that are applied on the same body that 
exerts the deformation, as long as no mean-motion resonance exists between the masses.

As a working example, we apply the model to the Kepler-38 binary system. The evolution of the 
spin equations shows that the planet reaches a stationary solution much faster than the stars, 
and the equilibrium spin frequency is sub-synchronous. The binary components, on the other 
hand, evolve on a longer timescale, reaching a super-synchronous solution very close to that 
derived for the 2-body problem. The orbital evolution is more complex. After reaching spin 
stationarity, the eccentricity is damped in all bodies and for all the parameters analyzed here. 
A similar effect is noted for the binary separation. The semimajor axis of the planet, on the 
other hand, may migrate inwards or outwards, depending on the masses and orbital parameters. In 
some cases the secular evolution of the system may also exhibit an alignment of the pericenters, 
requiring to include additional terms in the tidal model.

Finally, we derived analytical expressions for the variational equations of the orbital 
evolution and spin rates based on low-order elliptical expansions in the semimajor axis ratio 
$\alpha$ and the eccentricities. These are found to reduce to the well-known 2-body case when 
$\alpha \rightarrow 0$ or when one of the masses is taken equal to zero. This model allow us to 
find a close and simple analytical expression for the stationary spin rates of all the bodies, 
as well as predicting the direction and magnitude of the orbital migration. 
}

\maketitle

\section{Introduction}

As of 2019, the {\it {Kepler mission}} has discovered approximately ten circumbinary (CB) 
planetary systems. All binary components define compact systems with orbital periods less than 
$\sim 40$ days and a wide range of eccentricities and mass ratios. The planets surrounding them 
also have a diversity of masses (between super-Earths to Jupiter masses) but they are all almost 
coplanar with the binary. With the exception of Kepler 34b and Kepler 413b, all CB planets seem 
characterized by small semimajor axis low eccentricities.

While the low inclinations suggest that these planets formed in a CB disc aligned with the 
orbital plane of the central binary, it is well accepted that {\it{in situ}} formation so close 
to the binary is unlikely due to the strong eccentricity excitation induced by the secondary 
star \cite[e.g.][]{Lines2014,Meschiari2012}. However, as we move away from the binary, the 
gravitational potential approaches that of a single star and planetary formation appears to be 
easier, following usual core-accretion models. This suggest that CB planets could have formed 
farther out, later migrated inward due to interaction with a primordial disc and finally stalled 
near their current orbits by some mechanism (\cite{Dunhill2013}).

In a previous work \citep{Zoppetti2018}, we tested the possibility that the circumbinary 
planets may have halted its inward migration due to a capture in a high order mean-motion 
resonance (MMR) with the binary and, once the disc is dissipated, slowly escaped from the 
commensurability due to tidal forces. We applied this hypothesis to Kepler-38, a very old system 
in which captures in the 5/1 MMR had been reported with hydro-simulation \citep{Kley2014}. 
Tidal interactions were modeled following \cite{Mignard1979} and incorporated to a N-body 
integrator following the prescription detailed in \cite{Rodriguez2011}. We observed that while 
the binary orbit shrinks due to tidal interactions, the planet seemed to increase its semimajor 
axis, even after the system reached stationary solutions in the spin rates. We were unable to 
explain these findings, which in principle could have been caused by a non-consistent treatment 
of the tidal interactions between the different bodies of the dynamical system.

In this article we present and discuss a self-consistent tidal model for a multi-body system, 
in which all tidal forces between pairs are computed adopting a weak-friction (Mignard-type) 
model. While the model is general, we will focus primarily on the spin and orbital evolution 
of the CB planet. To allow for a simpler comparison with our previous results, we will once 
again employ Kepler-38 as a reference system \citep{Orosz2012}. However, we will also explore 
a wider range of system parameters as well as different initial orbital elements and spin 
rates. 

This paper is organized as follows. In Section \ref{model} we present the model in two steps: 
in Section \ref{sub1} we first discuss which tidal forces have a net effect onto the 
dynamical evolution of an 3-extended-body system while in Section \ref{sub2} we show how 
these forces are incorporated, self-consistently, into our tidal model. Section \ref{numsim}
presents a series of numerical integrations of the full spin and orbital equations of motion. We 
concentrate on two different time-scales: the early dynamical evolution of the system before the 
spins reached stationary solutions, and the subsequent long-term orbital evolution of the CB 
planet in spin stationarity. In Section \ref{anali}, we construct analytical expressions for the 
orbital and spin evolution, averaged over the orbital periods but retaining secular terms, 
including those containing the difference between longitudes of pericenter. These allow us to 
estimate the stationary spin rate of CB planets, as well as the direction and magnitude of the 
orbital migration. We compare these predictions with full N-body simulations. Finally, Section 
\ref{conclu} summarizes our main results and discusses their implications.

\section{The model}
\label{model}

Let us consider a binary system in which $m_0$ and $m_1$ are the masses of the stellar 
components and $m_2$ is a circumbinary planet. We suppose that all the bodies lie in the same 
orbital plane and their spins are perpendicular to it. We also assume that all the bodies are 
extended masses with physical radii ${\cal R}_i$ and are deformable due to tidal effects between 
them.

For the gravitational interactions between each pair of bodies we will be adopt the classical 
weak-friction tidal model \citep{Mignard1979}. However, since now the tidal deformation of 
each body will have to incorporate the gravitational potential generated by both of its 
companions, we first need to address two issues: (i) which tidal deformation have a net effect 
on the long-term dynamical evolution of the system and, (ii) how the different forces should be 
incorporated into a self-consistent physical model. These questions are addressed in the next 
two subsections.

\subsection{The Mignard forces revisited}
\label{sub1}

We begin considering our three-body system with two simplifications. First, we will neglect the 
gravitational perturbations generated by $m_2$ on the other two bodies, as well as the effects 
of $m_1$ on $m_2$. Second, only $m_0$ will be assumed to be an extended mass while $m_1$ and 
$m_2$ will be taken as point masses. As a consequence of these approximations, the dynamics 
of both $m_1$ and $m_2$ around $m_0$ will be defined by the point-mass approximation plus the 
tidal deformation of $m_0$ generated solely by $m_1$. The role of $m_2$ is thus reduced to serve 
as a tracker of the dynamical effect of the tidal bulge on any generic orbit in the 
configuration plane. 

\begin{figure}
\centering
\includegraphics[width=0.8\columnwidth,clip]{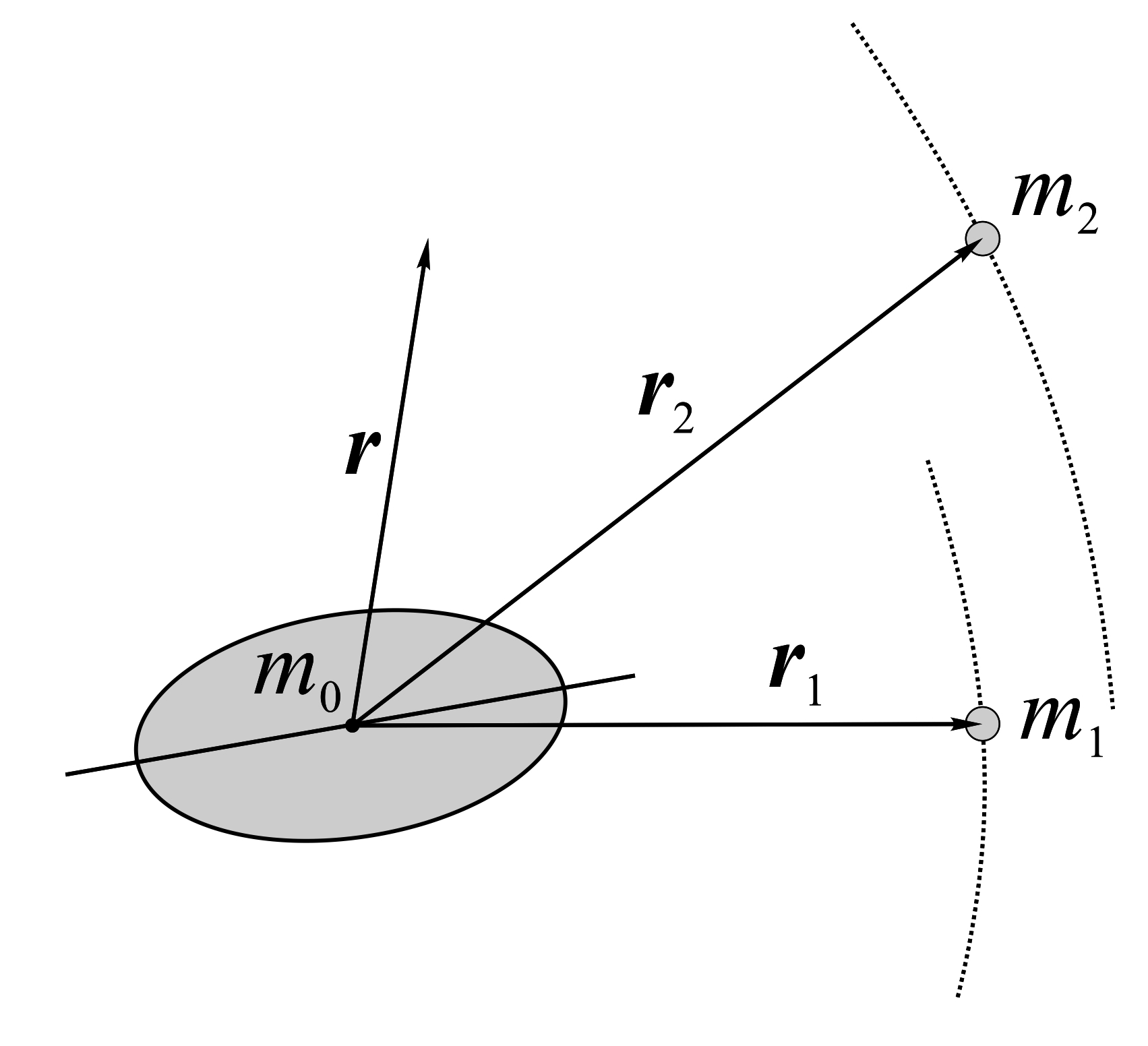}
\caption{A tidal lagged bulge generated on $m_0$ due to $m_1$ and its effect on  a test body 
$m_2$, from a $m_0$-centric coordinate frame.}
\label{fig:1}
\end{figure}

A schematics of this scenario is presented in Figure \ref{fig:1}, where ${\bf r_i}$ are the 
$m_0$-centric position vectors of the other masses. Following Mignard (1979), the tidal bulge 
of $m_0$ considered is displaced with respect to the instantaneous position of $m_1$ by a 
constant time-lag $\Delta t_0$. We assume the lag is sufficiently small to expand the 
gravitational potential $U$ generated by $m_0$ in anywhere in the space up to first-order in 
$\Delta t_0$, such that
\be
U({\bf {r}},{\bf {r}_1}) = U^{(0)}({\bf {r}},{\bf {r}_1}) + U^{(1)}({\bf {r}},{\bf {r}_1}) + 
\mathcal{O}(\Delta t_0^2)  
\label{eq1}
\ee
where $U^{(0)}$ and $U^{(1)}$ are the expanded tidal potentials of order $\mathcal{O}(0)$ and 
$\mathcal{O}(\Delta t_0)$, respectively. In particular, if we evaluate (\ref{eq1}) on the
position of $m_2$ (i.e. ${\bf {r}}={\bf {r}_2}$), we obtain 
\bea
\begin{split}
U^{(0)}({\bf {r}_2},{\bf {r}_1}) &= \frac{\mathcal{G} m_1 \mathcal{R}_0^5}{2 r_1^5 r_2^5} 
k_{2,0} \bigg[ 3 {( {\bf{r}_2} \cdot {\bf{r}_1})}^2 - r_2^2 r_1^2 \bigg]  \\
U^{(1)}({\bf {r}_2},{\bf {r}_1}) &= \frac{3 \mathcal{G} m_1 \mathcal{R}_0^5}{r_1^5 r_2^5} 
k_{2,0} \Delta t_0 \bigg[ \frac{({\bf{r_1}} \cdot \dot{{\bf{r}_1}})} {2 r_1^2} [5 
{({\bf{r}_2} \cdot {\bf{r}_1})}^2 - r_2^2 r_1^2 ] \\
& \hspace*{1.8cm} - ({\bf{r}_2} \cdot {\bf{r}_1}) [{\bf{r}_1} \cdot ({\bf{\Omega}}_0 \times 
{\bf{r}_2}) + {\bf{r}_2} \cdot \dot{{\bf{r}_1}}] \bigg] \\
\end{split}
\label{eq2}
\eea
where $\mathcal{G}$ is the gravitational constant, ${\bf{\Omega}}_0$ is the spin vector of 
$m_0$ and $k_{2,0}$ its the second degree Love number.

\begin{figure*}
\centering
\includegraphics[width=0.99\textwidth]{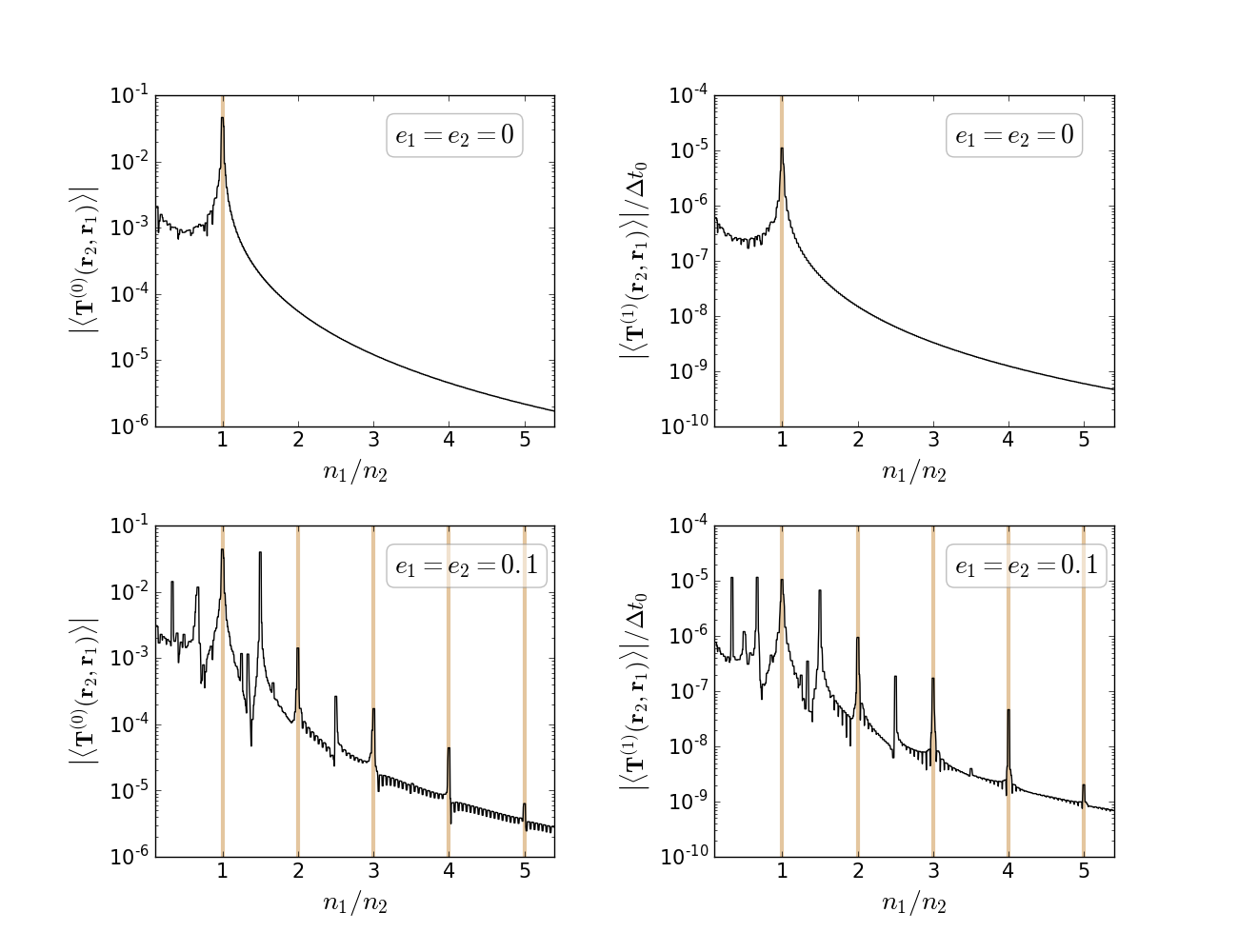}
\caption{Secular normalized torques of zero-order $|\langle {\bf{T}}^0({\bf{r}}_2) \rangle|$ 
(left column) and first-order $|\langle{\bf{T}}^{1}({\bf{r}}_2)\rangle|$ (right column),
computed on $m_2$ due to the tidal deformation on $m_0$ induced by $m_1$, plotted as a function 
of the mean-motion ratio $n_1/n_2$. We considered $m_0 = 1$, $a_1 = 1$ and varied $a_2$ to 
include orbits both interior and exterior to $m_1$. Upper panels correspond to circular orbits 
($e_1=e_2=0$) while the lower panels assume eccentric orbits with $e_1=e_2=0.1$. Light 
brown vertical lines highlight the location of some important mean-motion resonances. Note that 
the first-order torques in the right panels are also normalized respect to the time-lag $\Delta 
t_0$.}
\label{fig:torques}
\end{figure*}

The tidal force per unit mass ${\cal {\bf f}}$ generated by $m_0$ at a generic position vector 
${\bf r}$ can be obtained as 
\be
{\bf f} = \nabla_{\bf {r}} (U^{(0)} + U^{(1)}) = {\bf f}^{(0)} + {\bf f}^{(1)}
\label{eq3}
\ee
where explicit expressions evaluated on $m_2$ are given by
\bea
\begin{split}
{\bf f}^{(0)} &= \frac{3 \mathcal{G} m_1 \mathcal{R}_0^5}{2 
r_2^5 r_1^5} k_{2,0} \bigg[ 2 ( {\bf{r}_2} \cdot {\bf{r}_1}){\bf{r}_1}
+ \bigg({r_1}^2 -\frac{5}{r_2^2} {({\bf{r}_2} \cdot {\bf{r}_{1}})}^2  \bigg){\bf{r_2}} \bigg] 
\\
{\bf f}^{(1)} &= \frac{3 \mathcal{G} m_1 \mathcal{R}_0^5}{r_2^5 
r_1^5} k_{2,0} \Delta t_0 \bigg[ \frac{({\bf{r}_1} \cdot \dot{{\bf{r}_1}})}{r_1^2} [5 
{\bf{r}_1} ({\bf{r}_2} \cdot {\bf{r}_1}) - {\bf{r}_2} {r_1}^2 ] \\
&- [{\bf{r}}_1 \cdot ({\bf{\Omega}}_0 \times {\bf{r}_2}) + {\bf{r_2}} \cdot \dot{{\bf{r}_1}} ] 
{\bf{r}_1} - ({\bf{r}_1} \times {\bf{\Omega}}_0 + \dot{{\bf{r}_1}})({\bf{r}_2} \cdot 
{\bf{r}_1}) 
\\
&+ \frac{5 {\bf{r}_2}}{r_2^2} \bigg[({\bf{r}_2} \cdot {\bf{r}_1})[{\bf{r}_1} \cdot 
({\bf{\Omega}}_0 \times {\bf{r_2}}) + {\bf{r}_2} \cdot \dot{{\bf{r}_1}} ] \\
&- \frac{({\bf{r}_1} \cdot \dot{{\bf{r}_1}})}{2 r_1^2} [5 {({\bf{r}_2} \cdot {\bf{r}_{1}})}^2 
- {r}^2_{2} {r_1}^2 ] \bigg] \bigg] . \\
\end{split}
\label{eq4}
\eea
Finally, the torques per unit mass can be calculated as ${\bf{T}}({\bf {r}},{\bf {r}_1}) \simeq 
{\bf{r}} \times ( {\bf f}^{(0)}+{\bf f}^{(1)} ) = {\bf{T}}^{(0)}({\bf {r}},{\bf {r}_1}) + 
{\bf{T}}^{(1)}({\bf {r}},{\bf {r}_1})$. As before, evaluating on the position of $m_2$ yields
\bea
\begin{split}
{\bf{T}}^{(0)}({\bf {r}_2},{\bf {r}_1}) &= \frac{3 \mathcal{G} m_1 k_{2,0} 
\mathcal{R}_0^5}{r_2^5 
r_1^5} ({\bf{r}_2} \cdot {\bf{r}_1})({\bf{r}_2} \times {\bf{r}_1}) \\
{\bf{T}}^{(1)}({\bf {r}_2},{\bf {r}_1}) &= \frac{3 \mathcal{G} m_1 k_{2,0} 
\mathcal{R}_0^5}{r_2^5 r_1^5}\Delta t_0 \bigg[ 5 \frac{({\bf{r}_1} \cdot 
\dot{{\bf{r}}_1})}{r_1^2} ({\bf{r}_2} \cdot {\bf{r}_{1}}) 
({\bf{r}_{2}} \times {\bf{r}_1}) \\
&- [{\bf{r}_1} \cdot ({\bf{\Omega}_0} \times {\bf{r}_2}) + {\bf{r}_2} \cdot \dot{{\bf{r}}_1}] 
({\bf{r}_2} \times {\bf{r}_1}) \\
&- ({\bf{r}_2} \cdot {\bf{r}_1}) [ ({\bf{r}_{2}} \cdot {\bf{\Omega}}_0){\bf{r}_{1}} - 
({\bf{r}_{2}} \cdot {\bf{r}_1}){\bf{\Omega}_0} + {\bf{r}_2} \times \dot{{\bf{r}}_1} ] \bigg] 
.\\
\end{split}
\label{eq5}
\eea

In the classical two-body tidal problem, the acceleration ${\bf f}$ and the torque ${\bf {T}}$ 
are computed on the position of the deforming body $m_1$. It is easy to see that in such a case, 
the zero-order torque reduces to zero (i.e. ${\bf{T}}^{(0)} ({\bf{r}} = {\bf{r}_1},{\bf{r}_1}) = 
{\bf{0}}$) and the only net contribution to the orbital and spin evolution (notwithstanding a 
precession term) stems from the the first-order expressions ${\bf f}^{(1)}$ and ${\bf{T}}^{(1)}$ 
(see equations (5) and (6) of \cite{Mignard1979}). However, it is not immediately clear what 
occurs if ${\bf r} \ne {\bf r_1}$. In other words, we wish to analyze what are the (long-term) 
dynamical effects of a tidal bulge generated on $m_0$, due to the perturbing potential of $m_1$, 
on the orbit of another body $m_2$.

To address this question, let the $m_0$-centric orbits of $m_i$ be characterized by semimajor 
axis $a_i$, eccentricity $e_i$, mean longitude $\lambda_i$ and longitude of pericenter 
$\varpi_i$. Furthermore, let $n_i$ denote the mean-motion (orbital frequency) of each body.
We then compute the net secular torques ($\langle\bf{T}^{(0)}({\bf {r}}_2,{\bf{r}}_1)\rangle$ 
and $\langle\bf{T}^{(1)}({\bf {r}}_2,{\bf{r}}_1)\rangle$) for different values of $a_2$, 
assuming fixed values for $(a_1,e_1,e_2,\varpi_1,\varpi_2)$. The secular torques are calculated 
averaging over the short-period terms associated to $\lambda_1$ and $\lambda_2$. Since we will 
not restrict our analysis to non-resonant configurations between $m_1$ and $m_2$, we cannot 
assume that both mean longitudes are necessarily independent. We thus substitute the classical 
double averaging over $\lambda_i$ with a time average over time, such that
\begin{eqnarray}
\langle{\bf{T}}^{(i)}({\bf{r}_{2}},{\bf{r}_{1}})\rangle = \lim_{\tau \to \infty} 
\frac{1}{\tau} \int_{0}^\tau {\bf{T}}^{(i)} ({\bf{r}_2}(t),{\bf{r}_1}(t)) \, dt ,
\label{eq6}
\end{eqnarray}
with $i=0,1$. This technique allows us to evaluate the net secular contribution in both resonant 
and secular configurations of both bodies. In particular, the classical Mignard expressions 
should be obtained assuming equal orbits (and orbital frequencies) for $m_1$ and $m_2$. 

Results are shown in Figure \ref{fig:torques} for $m_0 = 1$, $a_1 = 1$, and $\varpi_1 = 
\varpi_2$ and $e_1=e_2$. The values of the averaged torques are normalized with respect to $m_1$ 
and $\Delta t_0$, and plotted as function of the mean-motion ratio $n_1/n_2$. The left-hand 
panels correspond to the zero-oder torque $\langle\bf{T}^{(0)}\rangle$, while the first-order 
contributions $\langle\bf{T}^{(1)}\rangle$ are depicted in the right-hand graphs.

In the upper panels we analyze the circular case ($e_1=e_2=0$) and in the lower panel eccentric 
orbits ($e_1=e_2=0.1$). Eccentricities are assume fixed throughout the numerical averaging. All 
plots show distinct peaks, where the net torque is different from zero, overlaid with respect 
to background values that decrease smoothly as $n_1/n_2 \rightarrow \infty$. This background 
trend is a consequence of the numerical approximation employed to evaluate the time integral 
(\ref{eq6}), which basically consisted in a discrete sum over a finite time interval equal to 
500 orbital periods of the outermost body. 

For circular orbits we observe that the only value of $a_2$ for which $m_2$ receives a non-zero 
net torque corresponds to a $1/1$ MMR, that is in a coorbital position with the deforming body 
$m_1$. In particular, the case in which the position of both masses coincide (i.e. 
$\lambda_1=\lambda_2$) yields results analogous to those obtained from the 2-body tidal model.
Different initial values of the mean longitudes would, in principle, allow us to estimate both 
torques in other coorbital configurations, such as that occurring for $m_2$ located in a 
Trojan-like orbit with the other masses. Finally, although $|\langle {\bf{T}}^0({\bf{r}}_2) 
\rangle|$ is different from zero in the $1/1$ MMR, its dynamical effect on the orbit reduces to 
a tidal precession term (e.g. Correia et al. 2011) and does not contribute to any secular 
changes in the orbits or spin rates.

The eccentric case, depicted in the lower frames, exhibits a richer diversity. Non-zero torques 
are found in several mean-motion resonances and not only in the coorbital region. This seems to 
imply that the tidal deformation generated by $m_1$ on $m_0$ should affect the dynamical 
evolution of $m_2$ whenever there exists a commensurability relation between the orbital 
frequencies. This is an important finding, indicating that tidal models for resonant bodies 
could require the full tidal deformation on each body as generated by all the other bodies 
of the system.

The numerical results described in Figure \ref{fig:torques} were confirmed introducing elliptic 
expansions for the position and velocity vectors in the equations (\ref{eq4}) and (\ref{eq5}), 
truncated up to fourth order in semimajor axis ratio and eccentricities, and integrating the 
resulting expressions analytically. 

In conclusion, in the absence of any mean-motion relation between $m_1$ and $m_2$, the only 
tidal forces that need to be considered on $m_i$ are those stemming from the deformation that 
$m_i$ generates on $m_j$ and $m_k$ (with $i \ne j \ne k$). Since the tidal deformation 
generated on $m_j$ by $m_k$ may be neglected, a multi-body tidal model may be constructed 
simply by adding the forces and torques between deformed-deforming pairs as given in equations 
(5) and (6) of \cite{Mignard1979}. The effect of the tidal torques on resonant orbits will be 
investigated in a forthcoming work. 

\begin{figure}
\centering
\includegraphics[width=0.9\columnwidth,clip=true]{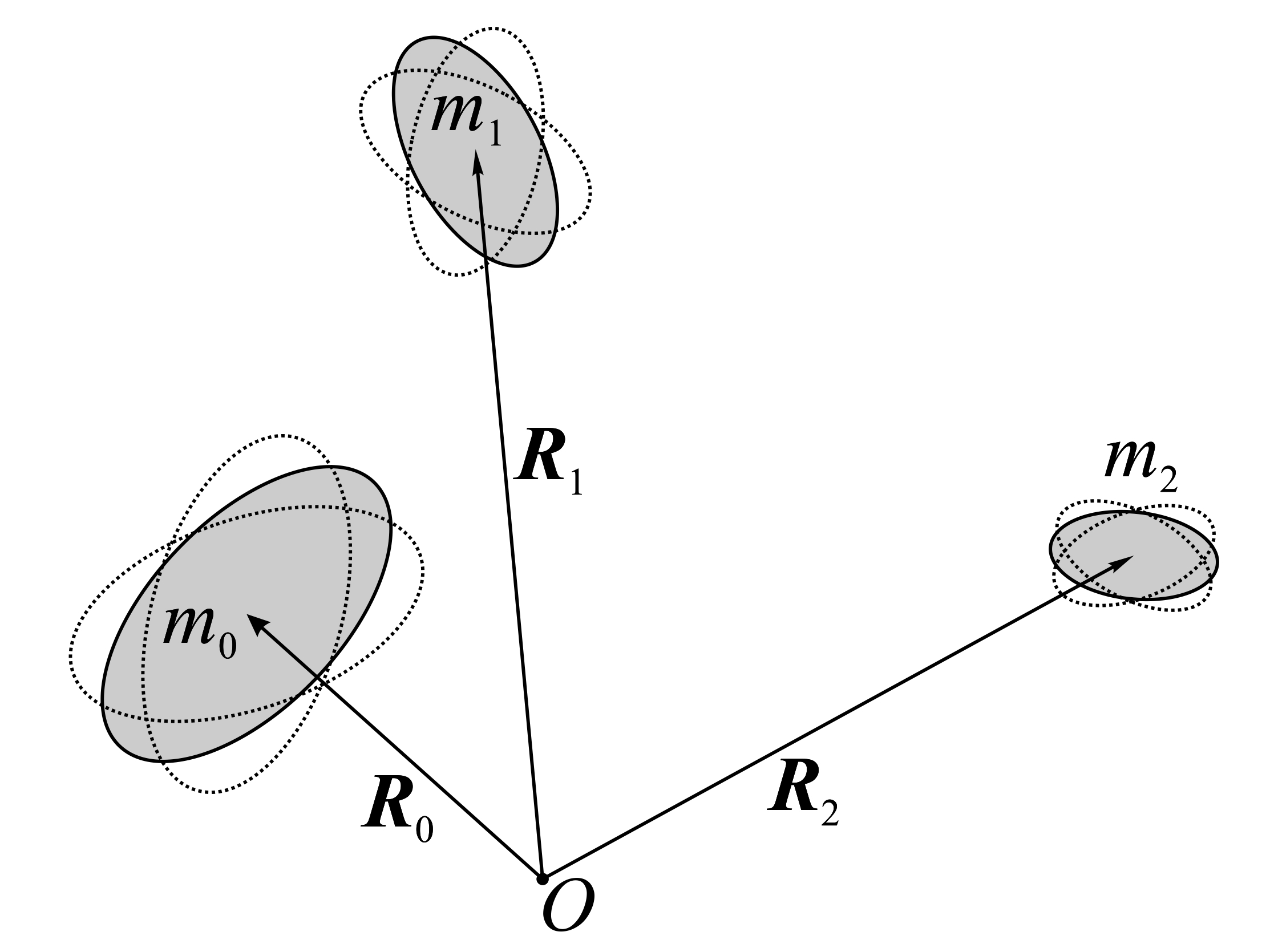}
\caption{Tidally interacting 3-body system. The tidal bulge generated on each body (filled 
gray ellipsoids) are the sum of the deformations generated by each of its companions. ${\bf 
R_i}$ denote the position vectors with respect to a generic inertial reference frame.}
\label{fig:3c}
\end{figure}

\subsection{The equations of motion}
\label{sub2}

Having identified the tidal forces affecting the long-term and secular dynamical evolution of 
the system, we now discuss how they should be incorporated into the equations of motion of the 
circumbinary system in a self-consistent manner. 

We return to our full circumbinary system where now all bodies are considered extended and 
gravitationally interacting. As shown in Figure \ref{fig:3c}, the equilibrium deformation of 
body $m_i$ is the sum of two ellipsoids, each generated by the gravitational potential of the 
other two masses. As shown in \cite{Folonier2017}, the sum of two ellipsoidal bulges can be 
approximated by a single ellipsoidal bulge with its own flattening and orientation. However, as 
discussed in Section \ref{sub1}, we only need to consider the direct distortion between pairs.

Let us denote by ${\bf R_i}$ the position vector of $m_i$ in an inertial reference frame.
Then, the complete equations of motion, including both tidal and point-mass terms, may be 
expressed as:
\bea
\begin{split}
m_0 {\bf \ddot{R}_0}  &= \;\;\, 
\frac{\mathcal{G}m_0 \, m_1}{|{\bf{\Delta_{10}}}|^3} {\bf \Delta_{10}}
+\frac{\mathcal{G}m_0 \, m_2}{|{\bf \Delta_{20}}|^3} {\bf \Delta_{20}} + \bf{F_0} \\
m_1 {\bf \ddot{R}_1}  &=  
-\frac{\mathcal{G}m_0 \, m_1}{|{\bf \Delta_{10}}|^3} {\bf \Delta_{10}}
+\frac{\mathcal{G}m_1 \, m_2}{|{\bf \Delta_{21}}|^3} {\bf \Delta_{21}} + \bf{F_1} \\
m_2 {\bf \ddot{R}_2}  &=  
-\frac{\mathcal{G}m_0 \, m_2}{|{\bf \Delta_{20}}|^3} {\bf \Delta_{20}}
-\frac{\mathcal{G}m_1 \, m_2}{|{\bf \Delta_{21}}|^3} {\bf \Delta_{21}} + \bf{F_2} 
\label{eq7}
\end{split}
\eea
where for compactness we have denoted the relative position vectors as
\be
{\bf \Delta_{ij}} \equiv {\bf R_i} - {\bf R_j} .
\label{eq8}
\ee
In terms of the $m_0$-centric position vectors, these are simply given by ${\bf \Delta_{10}} = 
{\bf r_{1}}$, ${\bf \Delta_{20}} = {\bf r_{2}}$ and ${\bf \Delta_{21}} = {\bf r_{2}} - {\bf 
r_{1}}$. The last terms of the equations of motion are the complete tidal forces acting on each 
mass. Following \cite{Ferraz-Mello2008}, considering the reacting forces, these may be 
expressed by
\bea
\begin{split}
\bf{F_0} &= \bf{F_{0,1}} + \bf{F_{0,2}} - \bf{F_{1,0}} - \bf{F_{2,0}} \\
\bf{F_1} &= \bf{F_{1,0}} + \bf{F_{1,2}} - \bf{F_{0,1}} - \bf{F_{2,1}} \\
\bf{F_2} &= \bf{F_{2,0}} + \bf{F_{2,1}} - \bf{F_{0,2}} - \bf{F_{1,2}} , \\ 
\end{split}
\label{eq9}
\eea
where ${\bf{F}_{i,j}}$ to the tidal force acting on $m_i$ due to the deformation in $m_j$. Note 
that the positive contributions in $\bf{F_i}$ are the direct effect of the deformation of the 
other bodies while the negative terms corresponds to the reaction of the force due to the 
deformation of $m_i$. These have the form
\bea
{\bf{F}_{i,j}} = -\frac{\mathcal{K}_{i,j}} {{|{\bf \Delta_{ij}}|}^{10}}
\bigg[ 2({\bf \Delta_{ij}} \cdot {\bf \dot{\Delta}_{ij}}) {\bf \Delta_{ij}} 
+ {\bf \Delta_{ij}}^2 ( {\bf \Delta_{ij}} \times {\bf{\Omega}_j} + {\bf \dot{\Delta}_{ij}} )  
\bigg] 
\label{eq10}
\eea
(Mignard 1979), where $\mathcal{K}_{i,j}$ is a measure of the magnitude of the tidal force and 
is given by
\be
\mathcal{K}_{i,j} = 3 \mathcal{G}m_i^2 \mathcal{R}_j^5 k_{2,j}\Delta t_j.
\label{eq11}
\ee
As before, ${\bf \Omega_j}$ is the spin angular velocity of $m_j$ and is assumed parallel to 
the orbital angular momentum. We have neglected the tidal contributions which arise from the 
zero-order potential since its effect is restricted to a precession of the pericenters and does 
not introduce any secular changes in the spins, semimajor axes or eccentricities.

\subsection{The rotational dynamics}

While the orbital dynamics can be obtained solving the equations of motion (\ref{eq7}), the 
time variation of the spins are deduced from the conservation of the total angular momentum
${\bf L_{\rm tot}}$. Since we assumed rotations perpendicular to the common orbital plane,
\be
{\bf L_{\rm tot}} = {\bf L_{\rm orb}} + \sum_{i=0}^2 C_i {\bf \Omega_i} = const.,
\label{eq12}
\ee
where $C_i$ is the principal moment of inertia of $m_i$. In turn, the orbital angular momentum 
in the inertial reference frame is given by
\be
{\bf L_{\rm orb}} = \sum_{i=0}^2 m_i ({\bf R_i} \times {\bf \dot{R}_i} ) . 
\label{eq13}
\ee

Differentiating this equation with respect to time and substituting expressions (\ref{eq7}) for 
the accelerations ${\bf \ddot{R}_i}$, we obtain
\bea
\begin{split}
{\bf \dot{L}_{\rm orb}} &= 
   {\bf \Delta_{10}} \times {\bf{F}_{1,0}} + {\bf \Delta_{20}} \times {\bf{F}_{2,0}} \\
&+ {\bf \Delta_{01}} \times {\bf{F}_{0,1}} + {\bf \Delta_{21}} \times {\bf{F}_{2,1}} \\
&+ {\bf \Delta_{02}} \times {\bf{F}_{0,2}} + {\bf \Delta_{12}} \times {\bf{F}_{1,2}} .\\
\end{split}
\label{eq14}
\eea
Furthermore, assuming that the variation in the spin angular momenta of the body $m_j$ is only 
due to the terms in ${\bf {\dot L}_{\rm orb}}$ associated to its deformation, we obtain
\be
C_j {\bf {\dot \Omega}_j} = - \sum_{i \ne j} {\bf \Delta_{ij}} \times {\bf{F}_{i,j}} .
\label{eq15}
\ee
Note than in the limit where the physical radius of $m_j$ reduces to zero (i.e. 
$\mathcal{R}_j=0$), the tidal terms ${\bf{F}_{i,j}}$ are also zero for all $i \ne j$, and 
equation (\ref{eq15}) is automatically satisfied. 

Finally, using expression (\ref{eq10}) for the tidal forces, the time evolution of the spin 
vectors are given by
\be
\frac{d{\bf \Omega_j}}{dt} = \frac{1}{C_j} \sum_{i \ne j} 
\frac{{\cal K}_{i,j}}{|{\bf \Delta_{ij}}|^6} \bigg[ \frac{{\bf \Delta_{ij}} \times {\bf 
{\dot \Delta}_{ij}}}{|{\bf \Delta_{ij}}|^2} - {\bf \Omega_j} \bigg] .
\label{eq16}
\ee
Contrary to the 2-body case (e.g. Ferraz-Mello et al. 2008), the time derivative of the spin 
is given by the sum of two distinct terms. Depending on the magnitudes of each tidal term, it 
is not immediately obvious what would be the equilibrium rotational frequencies associated to 
stationary solutions.

\section{Numerical simulations}
\label{numsim}

In order study the dynamical predictions of our model, we analyze the tidal evolution of a 
3-body system consisting of a single planet around a binary star. The orbital and rotational 
evolution will be followed solving the equations of motion (\ref{eq7}) for the orbit and 
equation (\ref{eq16}) for each of the spins.

As before, we choose the Kepler-38 system as a test case, previously studied in 
(\cite{Zoppetti2018}) using a simpler tidal model. Nominal values for system parameters and 
initial orbital elements are detailed in Table \ref{tab1}. Stellar masses and radii were taken 
from (\cite{Orosz2012}), while the value of $m_2$ was estimated from the semi-empirical 
mass-radius from (\cite{Mills2017}). The orbital elements of the secondary star respect to 
$m_0$ are those expected during the early stages of the system before tidal interactions had 
time to act (see \cite{Zoppetti2018}), assuming tidal parameters and moments of inertia equal 
to those given in the table. 

The orbital parameters for the planet are similar to those presented by \cite{Orosz2012}, while 
the value of $Q'_2$ is consistent with rocky bodies \citep{Ferraz-Mello2008}. However, it is 
important to stress that there is little dynamical constraint on the values of the planetary 
tidal parameters; the value adopted here is for illustrative purposes only. Finally, the 
parameters highlighted with an asterisks were varied in our different simulations.

We will focus our attention on two different timescales: (i) an early stage (up to $\sim 1-2$ 
Gyr) characterized by the evolution of the rotation rates towards stationary solutions, and 
(ii) the subsequent long-term dynamical orbital evolution of the system. In this second part we 
will concentrate primarily on the orbital migration and eccentricity damping of the planet.

\subsection{Early dynamical evolution}

\begin{figure}
\centering
\includegraphics[width=.99\columnwidth,clip=true]{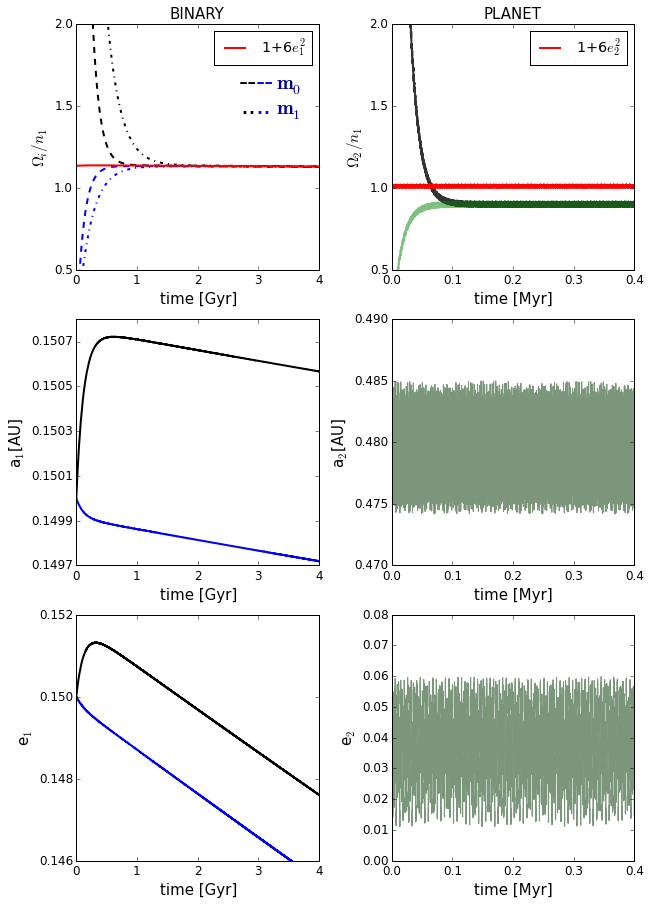}
\caption{Early tidal evolution of a circumbinary system. In all the panels, the black curve 
represents the results of our reference simulation (Table \ref{tab1}). {\bf Left:} Dynamical 
evolution of the binary, showing the spin rate (top), semimajor axis (middle) and eccentricity 
(bottom panel). The results depicted in blue consider initially slow-rotating stars with 
$\Omega_0/n_1=\Omega_1/n_1=0.1$ (at $t=0$), while those in black correspond to primordial fast 
rotators $\Omega_0/n_1=\Omega_1/n_1=10$. {\bf Right:} Evolution of the planetary spin and 
orbit. Black (respectively green) curves correspond to initial super-synchronous (respectively 
sub-synchronous) planetary spin rates. Time variation of the semimajor axis $a_2$ and 
eccentricity $e_2$ are practically equal in both cases (middle and bottom panels).}
\label{fig:shortevo}
\end{figure}

Figure \ref{fig:shortevo} shows the early rotational and orbital evolution of the binary stars 
and the planet. Except for the spin rates, all initial conditions and system parameters were 
taken equal to the nominal values of summarized in Table \ref{tab1}.

\begin{table}
\centering
\caption{Initial conditions for our reference numerical simulation, representing the primordial 
Kepler 38 system \protect\citep{Orosz2012, Zoppetti2018}. Orbital elements are given in a 
Jacobi reference frame. The parameters highlighted with an asterisk were varied in different 
simulations as indicated in the text.}
\label{tab1}
\begin{tabular}{lccc} 
\hline
body & $m_0$ & $m_1$ & $m_2$ \\
\hline
mass                       & $0.949 \, M_\odot$ & $0.249 \, M_\odot$   & $10 \, M_\oplus$   \\
radius                     & $0.84 \, R_\odot$  & $0.272 \, R_\odot$   & $4.35 \, R_\oplus$ \\
$C_i/(m_i\mathcal{R}_i^2)$ & $0.07$             & $0.25$               & $0.25$             \\
$Q'_i$                     & $1 \times 10^6$    & $1 \times 10^6$      & $1 \times 10^1$(*) \\
$\Omega_i$                 & $10 \, n_1$(*)     & $10 \, n_1$(*)       & $10 \, n_2$(*)     \\
$a_i$ [AU]                 &                    & $0.15$               & $0.48$             \\
$e_i$                      &                    & $0.15$               & $0.05$(*)          \\
\hline
\end{tabular}
\end{table}

We begin our analysis with the binary components, shown in the left-hand plots of Figure 
\ref{fig:shortevo}. The blue curves correspond to initial spin rates for both stellar 
components equal to $\Omega_0 = \Omega_1 = n_1/10$ (i.e. slow rotators), while the black
curves show results where the star were considered initially fast rotators: $\Omega_0 = \Omega_1 
= 10 n_1$. Regardless of the initial spin, both stars reach a pseudo-synchronization state in a 
few Gyrs, with a final rotational frequency equal to the value predicted by 2-body tidal models:
$\Omega_0/n_1=\Omega_1/n_1=1+6 e_1^2$ (e.g. Ferraz-Mello et al. 2008). 

If the stars were initially super-synchronous, the change in spin rates deliver angular momenta 
to their orbit (eq. \ref{eq15}), increasing the semimajor axis $a_1$ and eccentricity 
$e_1$. The opposite effect is observed if the stars were initially sub-synchronous: the orbit 
delivers angular momenta to the stars to increase their spins, decreasing its semimajor axis and 
eccentricity. Once the rotational stationary solution is attained, the subsequent dynamical 
effect of the stellar tides acts to reduce the semimajor axis and damp the eccentricity until 
the circularization is reached (\cite{Correia2016}, \cite{Hut1980}). Due to its small mass, the 
presence of the CB planet has no noticeable influence on the tidal evolution of the binary.

The right-hand panels of Figure \ref{fig:shortevo} show the dynamical evolution of the 
planetary spin (top panel) and orbit (middle and bottom plots). As before we considered two 
different initial spin rates: $\Omega_2/n_2 = 10$ is shown in black while $\Omega_2/n_2 = 0.1$ 
in green. The stellar spins were taken equal to the nominal values. We found no appreciable 
change in the time evolution of the semimajor axis or eccentricity regardless of the initial 
spins and, as seen in the middle and lower panels, both curves are practically 
indistinguishable. 

Concerning the evolution of the planetary spin, both initial conditions reach stationary values 
much faster than the stars (typically in a few Myrs), although the equilibrium value is 
sub-synchronous and significantly displaced with respect to the 2-body expectation (red 
horizontal line). This behavior will be discussed in detail in section \ref{anali_spin} and 
constitutes a new finding. Instead of the super-synchronous stationary solutions found in 
classical tidal models for eccentric orbits, the interacting binary system leads to a stable 
sub-synchronous state which does not change even after the stars themselves evolve towards their 
rotational stationary spins.

\begin{figure*}
\centering
\includegraphics[width=.8\textwidth,clip=true]{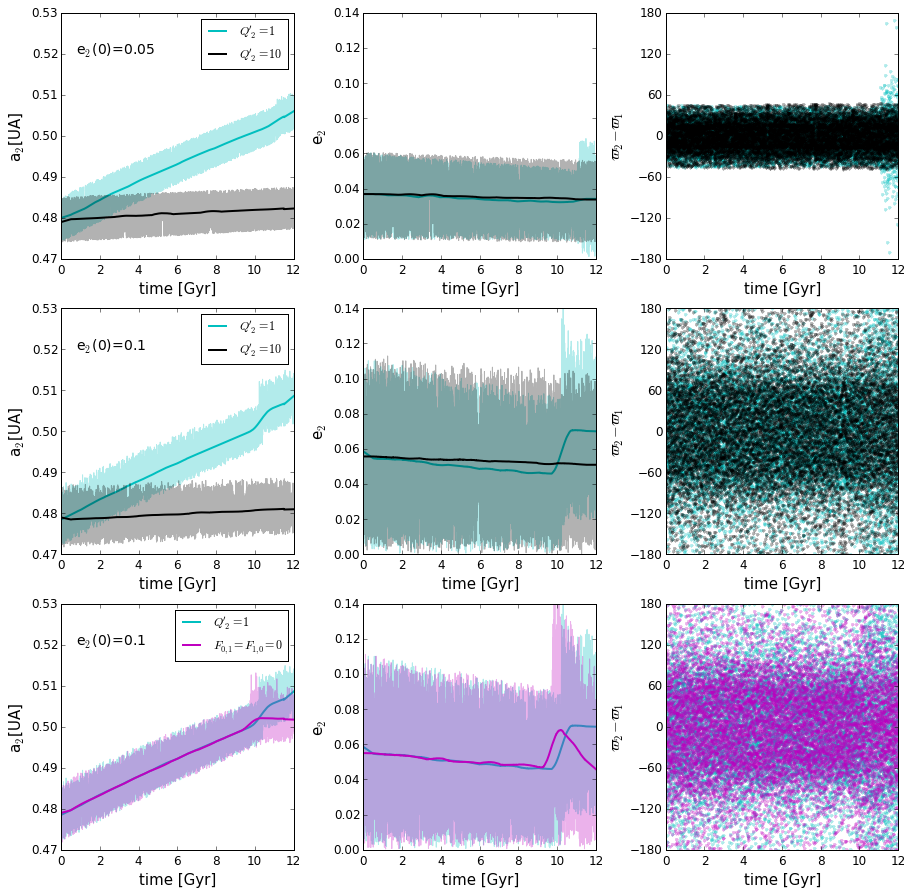}
\caption{Long-term orbital tidal evolution of our the planet in our Kepler-38-like system. 
Except for the parameters inlaid in the left-hand plots, all parameters and initial 
conditions were taken equal to those in Table \ref{tab1}. Light-tone curves for $a_2$ and 
$e_2$ show osculating values while darker curves correspond to mean elements obtained from a 
digital filter. The magenta curves in the lower panels are the result of a simulation 
disregarding tidal interaction between the stars.}
\label{fig:planetevo}
\end{figure*}

\subsection{Long-term orbital evolution}

Figure \ref{fig:planetevo} shows three different long-term simulations, integrated over 
timescales comparable with the estimated age of Kepler-38 system (\cite{Zoppetti2018}). All 
system parameters and initial conditions were chosen equal to their nominal values (Table 
\ref{tab1}) except for those described in the left-hand panels of each set. In all cases the 
planetary spin reached a sub-synchronous stationary solution early in the simulation; thus 
we concentrate on the orbital elements: semimajor axis $a_2$ in the left-hand plots, 
eccentricity $e_2$ in the center graphs, and difference between longitudes of pericenter 
$\Delta \varpi = \varpi_2 - \varpi_1$ in the right-hand graphs. Results after the application 
of the low-pass filter are shown in darker curves for $a_2$ and $e_2$.

The black curves in the top panels correspond to the results of our reference simulation (see 
Table \ref{tab1}) while in the cyan curves we consider a more dissipative planet with $Q'_2=1$. 
The middle panels show results considering a more eccentric initial orbit $e_2(0)=0.1$, again 
for the same two values of the tidal parameter. Finally, in the lower panel we analyze the case 
in which the stars in the binary are not tidally interacting. This scenario correspond to 
setting ${\bf{F}_{0,1}}={\bf{F}_{1,0}}=0$ (see eq. \ref{eq10}) in our code. Results with 
non-tidally interacting stars are shown in magenta, while cyan curves repeat the results of our 
simulation with tidal effects for the stars. 

Independently of the adopted tidal parameter $Q'_2$, the planet is always observed to migrate
outwards, marking a second distinct difference with respect to expectations from classical 
2-body tidal models. This result was already described in \cite{Zoppetti2018}, although in that 
case we used a simpler and non-consistent tidal model. Lower values of $Q'_2$ (cyan curves 
in the upper and middle panels) lead to more larger excursions in semimajor axis, ultimately 
leading to scattering in a high-order MMR and temporary excitation of the eccentricity. The 
magenta curve in the lower panels show that the outwards migration is not a consequence of 
tidal effects in the stars, but seems to be independent of their tidal evolution.

The planetary eccentricity, on the other hand, always seems to decrease, as long as not 
mean-motion resonances are encountered. For low initial values of $e_2$ (upper panels of 
Figure \ref{fig:planetevo}) the planet and secondary star enter an aligned secular mode 
(\cite{Michtchenko2004}) in which $\Delta \varpi$ librates around zero. The amplitude of 
oscillation increases for larger initial eccentricities until $\Delta \varpi$ is observed to 
circulate for $e_2(0)=0.1$. However, the libration/circulation is purely kinematic and the 
difference in behavior is related to the amplitude of oscillation of the eccentricity. 
Regardless, these results seem to indicate that an analytical model for the tidal evolution of 
these type of systems must include terms involving the secular angle $\Delta \varpi$, even if 
the tidal evolution timescales are much longer than those associated to the precession of 
pericenters $\varpi_1$ and $\varpi_2$.

\section{Analytical secular model}
\label{anali}

In order to construct an analtical model from the equations of motion (\ref{eq7}) and 
(\ref{eq16}), we first introduce a Jacobi reference frame for the position and velocity vectors 
of the bodies. In terms of the inertial coordinates ${\bf R_i}$, the positions of the 
masses in Jacobi coordinates are given by:
\bea
\begin{split}
{\boldsymbol \rho_{\bf 0}} &= \frac{1}{\sigma_2} (m_0 \, {\bf R_0} + m_1 \, {\bf R_1} + m_2 \, 
{\bf R_2}) \\
{\boldsymbol \rho_{\bf 1}} &= {\bf R_1} - {\bf R_0}  \\
{\boldsymbol \rho_{\bf 2}} &= {\bf R_2}  - \frac{1}{\sigma_1} (m_0 \, {\bf R_0} + m_1 \, {\bf 
R_1}) ,\\
\end{split}
\label{eq17}
\eea
where
\be
\sigma_i=\sum_{k=0}^i m_k .
\label{eq18}
\ee
Analogous expressions relate the velocities vectors in both reference systems.

\subsection{Secular evolution of the planetary spin}
\label{anali_spin}

Expanding the position and velocity vectors in equation (\ref{eq16}) up to second order in 
$\alpha = a_1/a_2$ and the eccentricities, and averaging with respect to both mean longitudes, 
we finally obtain the rate of change of the rotational frequency of the planet as:
\be
\left<\frac{d\Omega_2}{dt}\right> = \frac{1}{2 C_2 a_2^6} 
\sum_{i,j,k=0}^2 A^{(s)}_{i,j,k} \, \alpha^i e_1^j e_2^k ,
\label{eq19}
\ee
where the non-zero coefficients different are given by
\bea
\begin{split}
A^{(s)}_{0,0,0} &= 2 ({\cal K}_{0,2}+{\cal K}_{1,2}) (n_2-\Omega_2) \\
A^{(s)}_{2,0,0} &= 6(\gamma_0^2{\cal K}_{0,2}+\gamma_1^2 {\cal K}_{1,2}) (4n_2-n_1-3\Omega_2)\\ 
A^{(s)}_{2,2,0} &= 3(\gamma_0^2{\cal K}_{0,2}+\gamma_1^2 {\cal K}_{1,2})(12n_2+n_1-9\Omega_2)\\
A^{(s)}_{0,0,2} &= 3 ({\cal K}_{0,2}+{\cal K}_{1,2}) (9 n_2-5\Omega_2) \\
A^{(s)}_{2,0,2} &= 12 (\gamma_0^2 {\cal K}_{0,2}+\gamma_1^2 {\cal K}_{1,2}) (44 n_2 - 7n_1 - 
21\Omega_2) \\
A^{(s)}_{1,1,1} &= 9 (8 n_2 - 5 \Omega_2) (\gamma_0 {\cal K}_{0,2}+\gamma_1 
{\cal K}_{1,2}) \cos(\Delta\varpi) ,\\
\end{split}
\label{eq20}
\eea
with
\be
\gamma_0 =  \frac{m_1}{\sigma_1} \hspace*{0.4cm} ; \hspace*{0.4cm} 
\gamma_1 = -\frac{m_0}{\sigma_1} .
\label{eq21}
\ee

The stationary spin rate $\big< \Omega_2 \big>_{\rm stat}$ predicted by this equation can be 
easily calculated by equating expression (\ref{eq19}) to zero. The explicit form of the 
equilibrium rotational frequency was found to be 
\bea
\begin{split}
\big< \Omega_2 \big>_{\rm stat} &= (1 + 6 e_2^2) n_2 - 6 \frac{\gamma_0^2 
\gamma_1^2}{\gamma_0^2 + \gamma_1^2} \left( n_1-n_2 \right) \alpha^2 \\
&+ 3 \frac{\gamma_0^2 \gamma_1^2}{\gamma_0^2 + \gamma_1^2} \left( n_1 + 3n_2 \right) \alpha^2  
e_1^2  \\
&- 3 \frac{\gamma_0^2 \gamma_1^2}{\gamma_0^2 + \gamma_1^2} \left( 13 n_1 - 41 n_2 \right) 
\alpha^2 e_2^2 \\
&+ \frac{27}{2} \frac{\gamma_0 \gamma_1 (\gamma_0+\gamma_1)}{\gamma_0^2 + \gamma_1^2} n_2
\alpha e_1 e_2 \cos(\Delta \varpi) . \\
\end{split}
\label{eq22}
\eea
In the limit case in which the mass of one of the stars reduces to zero we recover the classical 
2-body super-synchronous stationary solution $\big<\Omega_2 \big>_{\rm stat} = (1+6 e_2^2) 
\, n_2$ (\cite{Ferraz-Mello2008},\cite{Correia2011}). On the other hand, we can observe that 
for low binary and planetary eccentricities ($e_1,e_2 \to 0$), the CB planet stationary solution 
is sub-synchronous by a factor that decreases proportional to $\alpha^2$ as we move outward 
from the binary, and is maximum for equal-mass stars $m_1=m_0$.

\begin{figure}
\centering
\includegraphics[width=1.05\columnwidth,clip=true]{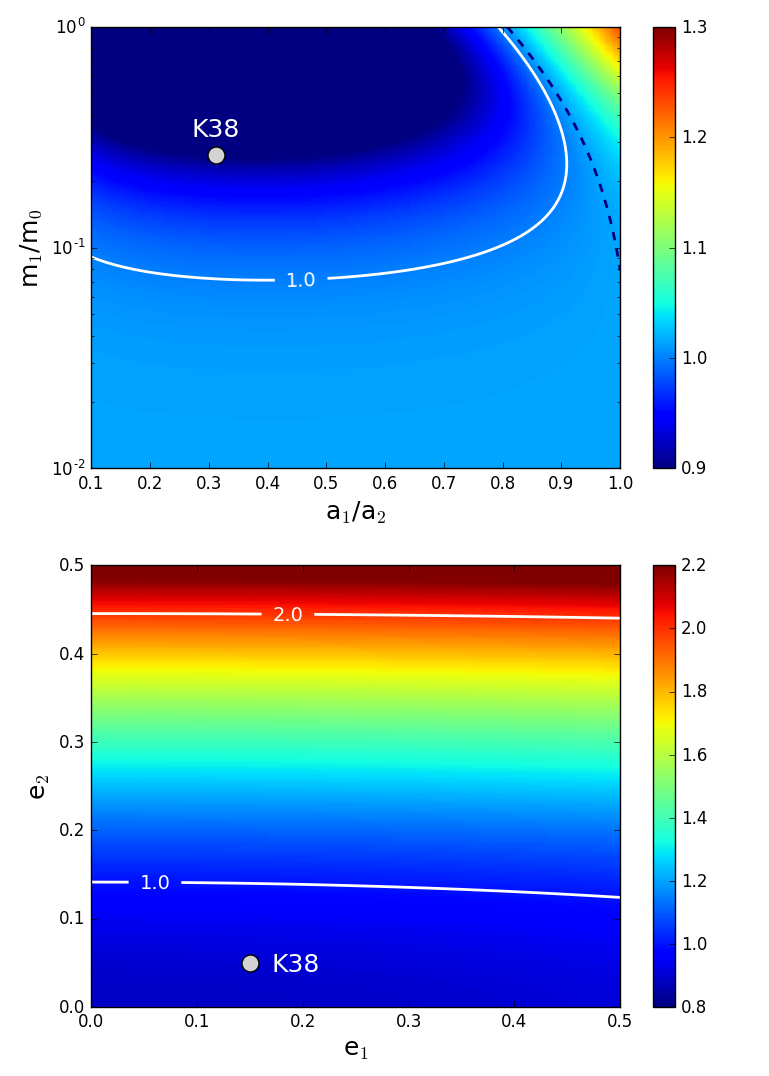}
\caption{Stationary planetary spin $\big<\Omega_2 \big>_{\rm stat}/n_2$ as function of the 
semimajor axis ratio $\alpha$ and mass of the secondary star (top), and as function of the 
eccentricities (bottom). All other system parameters were chosen equal to those given in 
Table \ref{tab1}. The nominal parameters for Kepler-38 are highlighted with a filled white 
circle and marked as ``K38''. Dashed curve in the top graph corresponds to $\big<\Omega_2 
\big>_{\rm stat}/n_2 = 1 + 6 e_2^2$.}
\label{fig:stat}
\end{figure}

Figure \ref{fig:stat} shows two color plots with the value of $\big<\Omega_2 \big>_{\rm stat}$ 
as a function of different system parameters and eccentricities (assumed constant). The top 
frame shows the dependence of the equilibrium spin rate of the planet with the distance from 
the binary system and the mass of the secondary star. Except for initial conditions very 
close to the binary or $m_1/m_0 \lesssim 0.1$, the estimated value of $\big<\Omega_2 \big>_{\rm 
stat}$ is always sub-synchronous with respect to the mean orbital frequency. The dashed black 
curve corresponds to the equilibrium value of the spin as obtained from the 2-body problem, 
i.e. $\big<\Omega_2 \big>_{\rm stat}/n_2 = 1 + 6 e_2^2$. Our model predicts lower values for 
practically all values of the system parameters, at least for the nominal eccentricities. This 
seems to imply that even a low-mass secondary, or even a large interior planet may counteract 
the super-synchronous state deduced from the 2-body solution and lead to appreciable 
differences in the rotational dynamics.

The dependence of $\big<\Omega_2 \big>_{\rm stat}$ with the eccentricities is analyzed in the 
bottom frame of Figure \ref{fig:stat}. We note that the sub-synchronous equilibrium state is 
only observed for low eccentricities of the planet, typically $e_2 \lesssim 0.1-0.15$, while 
super-synchronous states may be attained form more eccentric planets. However, since we expect 
tidal effects to damp the eccentricity, it appears that $\big<\Omega_2 \big>_{\rm stat} < n_2$ 
should probably the most common configuration in real-life systems. Finally, we observe little 
sensitivity of the equilibrium spin with respect to the eccentricity of the binary.

\begin{figure}
\centering
\includegraphics[width=1.07\columnwidth,clip=true]{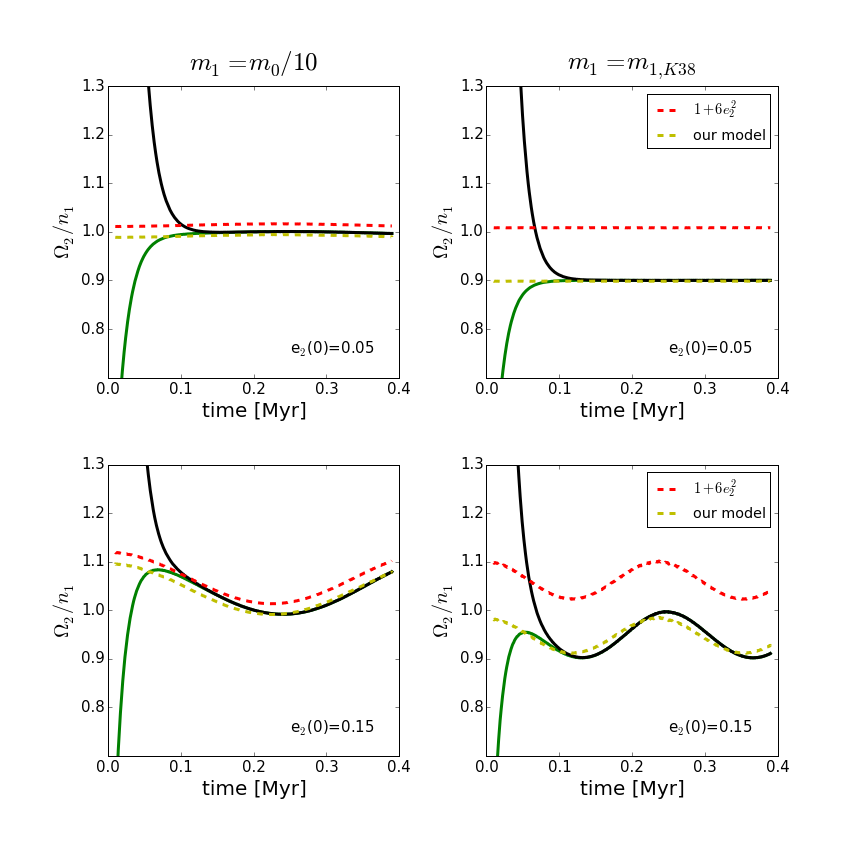}
\caption{N-body simulation of the spin evolution of fictitious CB planets, considering binaries 
with different mass ratios (different columns) and different initial eccentricity for the 
planets (different rows). In all the panels, the black curves correspond to the evolution of an 
initially super-synchronous planet while the green curves represent the initially 
sub-synchronous case. The dashed yellow curves are the stationary spins predicted by our model 
(eq. \ref{eq22}) while the dashed red curves are the 2-body stationary solution.}
\label{fig:subsin}
\end{figure}

In order to test the validity and precision of our analytical model, Figure \ref{fig:subsin} 
shows four sets of different N-body simulations of the evolution of the planetary spin, 
considering binaries with different mass ratios and planets in orbits with different initial
eccentricities. All results were digitally filtered to remove short-period variations. 

The top right-hand panel uses initial conditions from Table \ref{tab1} while the bottom 
right-hand panel considers a more eccentric CB planet. The left panels explore the case in 
which the mass of the secondary star is smaller than the nominal value. In every case the black 
curves correspond to initially super-synchronous planets while the green curves correspond to 
initially sub-synchronous CB planets. In dashed yellow curve, we show the synchronization spin 
predicted by our model (eq. \ref{eq22}) while in dashed red curve we compare with the 
stationary 2-body solution.

In accordance with the initial simulations presented in the previous section, the planetary 
spin reaches a stationary state rapidly, typically in about $10^5$ years, and our model 
seems to reproduce the equilibrium behavior extremely well. In the case of low-massive 
secondary star (left panels), the synchronization spin is very close to that predicted by the 
2-body model. However, when we consider binaries with mass ratios similar to Kepler-38 system, 
the synchronization spin is very different: sub-synchronous by an amount that can be very large 
for binaries with stars of comparable mass. Since the gravitational interaction causes 
long-term (secular) variations in the eccentricity of the planet, the value of $\Omega_2$ also 
suffers periodic oscillations. 

Finally, as can be observed from equation (\ref{eq22}), the stationary spin solution for the 
CB planets is not a function of the planetary mass $m_2$ nor of the physical radii of the 
bodies. Thus, if we assume that all currently known circumbinary planets have reached their 
stationary spin, we can predict their current rotational period just from the stellar masses and 
planetary orbits. As an example, considering its maximum possible eccentricity 
\citep{Orosz2012} and that the planet is in an aligned secular mode \citep{Zoppetti2018}, we 
estimate the rotation period of the planet in the Kepler-38 system in $P_{K38} \simeq 118$ days, 
about a $12 \%$ higher than the one predicted by the 2-body synchronization model.

\subsection{Variational equations for the orbital evolution}

Having developed an analytical model for the rotational dynamics, we turn our attention to the 
time evolution of the semimajor axis $a_2$ and eccentricity $e_2$. As before, we will focus on 
the planetary orbit, although analogous expressions can be found also for the binary. 

Following \cite{Beutler2005}, the variational equation for the semimajor axis in the 
Jacobi reference frame may be written as
\be
\frac{da_2}{dt} = \frac{2 a^2_2 }{\mathcal{G}\sigma_2} (\dot{\boldsymbol{\rho}}_{\bf 2} \cdot 
\delta\bf{f_2}) 
\label{eq23}
\ee
where $\delta{\bf{f}_2}$ is the total tidal force (per unit mass) affecting the 2-body motion 
of the planet around the center of mass of $m_0$ and $m_1$, and has the form:
\be
\delta{\bf{f}_2} = \frac{{\bf{F}_2}}{m_2}-\frac{1}{\sigma_1}({\bf{F}_0}+{\bf{F}_1}) .
\ee
Substituting equation (\ref{eq9}) in order to express the total force in terms of the 
individual two-body tidal interactions, we obtain
\be
\delta{\bf{f}_2} = \frac{1}{\beta_2} \bigg[ ({\bf{F}_{2,0}} - {\bf{F}_{0,2}}) + ({\bf{F}_{2,1}} 
- {\bf{F}_{1,2}}) \bigg] ,
\label{eq25}
\ee
where
\be
\beta_i = \frac{m_i\sigma_{i-1}}{\sigma_i}
\label{eq26}
\ee
is the reduced-mass \cite[e.g.][]{Beauge2007}. An analogous reasoning leads to a similar 
equation for the binary orbital evolution.

Expression (\ref{eq25}) shows that the total tidal force $\delta{\bf{f}}_2$ may be written in 
terms of differences of the type $({\bf{F}_{2,j}} - {\bf{F}_{j,2}})$, where $j=0,1$. From 
equations (\ref{eq10}), each of these differences may be explicitly written as
\bea
\begin{split}
{\bf{F}_{2,j}}-{\bf{F}_{j,2}} &= -\frac{\mathcal{K}^{(+)}_j}{|{\bf{\Delta_{2j}}}|^{10}} 
 \bigg[ 2({\bf \Delta_{2j}} \cdot {\bf \dot{\Delta}_{2j}}) {\bf \Delta_{2j}} \\
&\hspace*{2cm} + {\bf \Delta_{2j}}^2 ( {\bf \Delta_{2j}} \times {\bf{\bar {\Omega}}^{(j)}_2} 
+ {\bf \dot{\Delta}_{2j}} ) \bigg] \\
\end{split}
\label{eq27}
\eea
where we have defined
\be
{\cal K}_{j}^{(+)}={\cal K}_{2,j}+{\cal K}_{j,2}
\label{eq28}
\ee
and a new ``averaged'' rotational frequency
\be
{\bf {\bar {\Omega}}^{(j)}_2} = 
\frac{\mathcal{K}_{2,j}{\bf{\Omega}_j}+\mathcal{K}_{j,2}{\bf{\Omega}_2}}{{\cal K}_{2,j}+{\cal 
K}_{j,2}}.
\label{eq29}
\ee
Notice that expression (\ref{eq27}) has the same functional form as the tidal force in the 
2-body problem (eq. \ref{eq10}) with a magnitude given by ${\cal K}_{j}^{(+)}$ and a 
rotational frequency defined by ${\bf{\bar {\Omega}}^{(j)}_2}$. In the limit where $m_1 
\rightarrow 0$ and ${\cal R}_1 \rightarrow 0$, the term in the tidal force associated to ${\cal 
K}^{(+)}_1$ becomes negligible and we recover the same expression as found in the classical 
2-body case.

Writing the tidal forces in terms of Jacobi coordinates through ${\bf \Delta_{2j}} = 
\boldsymbol{\rho}_2 + \gamma_j \boldsymbol{\rho}_1$, substituting in the Gauss equation 
(\ref{eq23}), expanding in power series of $\alpha$, $e_1$ and $e_2$ and, finally, averaging 
over the mean longitudes, we obtain:
\be
\bigg< \frac{da_2}{dt} \bigg> = \frac{n_2}{ {\cal G} m_2 \sigma_2 a_2^4} 
\sum_{i=0}^4 \sum_{j,k=0}^2 \sum_{l=0}^1 A^{(a)}_{i,j,k,l} \, {\cal K}_{l}^{(+)} 
\gamma_l^i \alpha^i e_1^j e_2^k ,
\label{eq30}
\ee
where the non-zero coefficients are explicitly given by
\bea
\begin{split}
A^{(a)}_{0,0,0,l} &=   2 \Big[    \bar{\Omega}^{(l)}_2 - n_2                \Big] \\
A^{(a)}_{2,0,0,l} &=   2 \Big[ 12 \bar{\Omega}^{(l)}_2 + 5 n_1 - 17 n_2     \Big] \\
A^{(a)}_{4,0,0,l} &=  20 \Big[  6 \bar{\Omega}^{(l)}_2 + 5 n_1 - 11 n_2     \Big] \\
A^{(a)}_{2,2,0,l} &=     \Big[ 36 \bar{\Omega}^{(l)}_2 - 5 n_1 - 51 n_2     \Big] \\
A^{(a)}_{4,2,0,l} &= 100 \Big[  6 \bar{\Omega}^{(l)}_2 + n_1   - 11 n_2 \Big] \\
A^{(a)}_{1,1,1,l} &=   6 \Big[ 12 \bar{\Omega}^{(l)}_2 - 19 n_2 \Big] \cos(\Delta\varpi) \\
A^{(a)}_{3,1,1,l} &= \frac{25}{2} \Big[ 96 \bar{\Omega}^{(l)}_2 + 32 n_1 - 193 n_2 \Big] 
\cos(\Delta\varpi) \\
A^{(a)}_{0,0,2,l} &=     \Big[ 27 \bar{\Omega}^{(l)}_2 - 46 n_2 \Big] \\
A^{(a)}_{2,0,2,l} &=     \Big[ 528 \bar{\Omega}^{(l)}_2 + 5 (44 n_1 - 227 n_2) \Big] \\
A^{(a)}_{4,0,2,l} &=  10 \Big[ 390 \bar{\Omega}^{(l)}_2 + (325 n_1 - 1008 n_2) \Big] .\\
\end{split}
\label{eq31}
\end{eqnarray}

\begin{figure}
\centering
\includegraphics[width=1.0\columnwidth,clip=true]{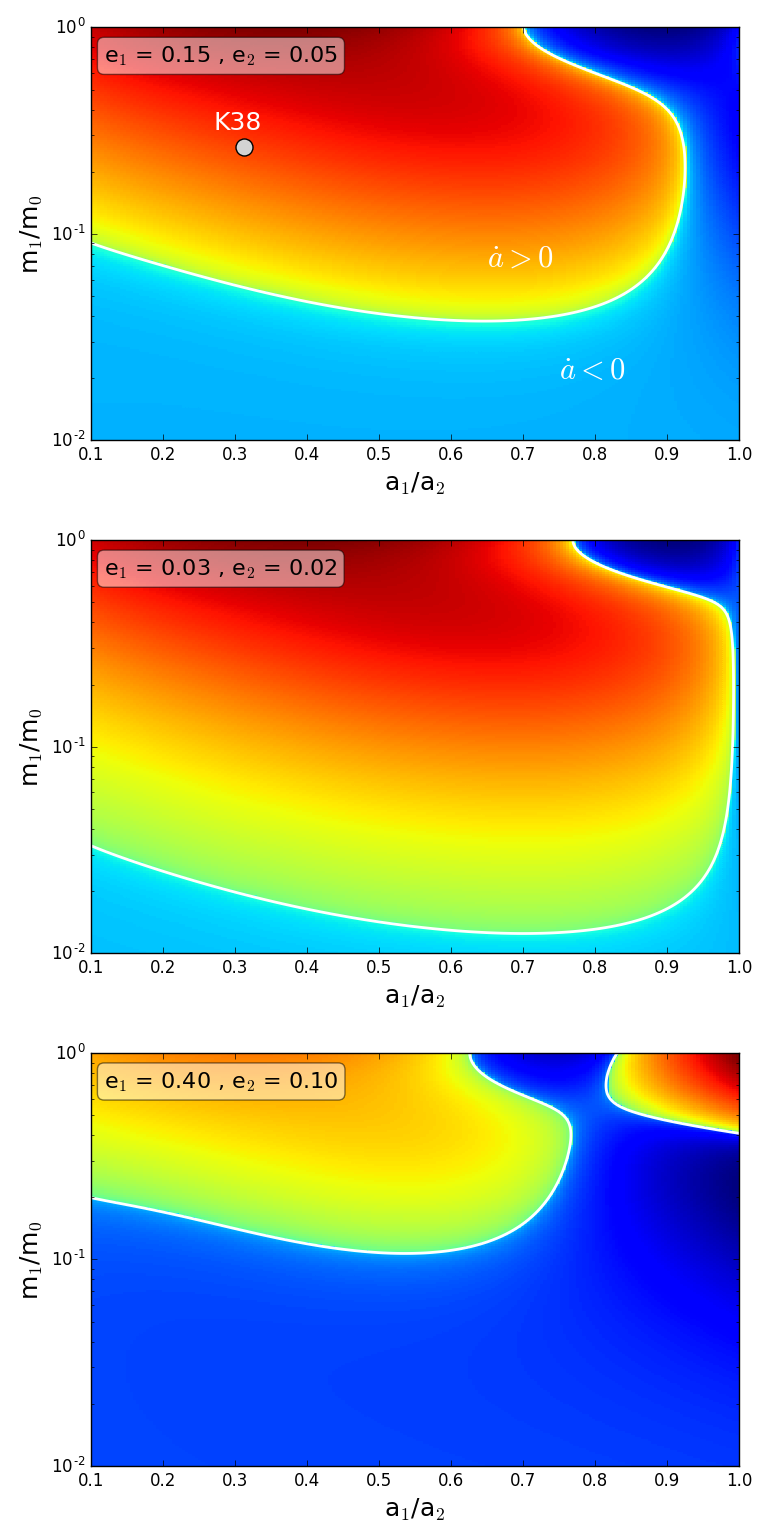}
\caption{Normalized values of the secular rate of change of the planetary semimajor axis, as 
function of the binary mass ratio and $\alpha$. Each panel shows results for different 
eccentricities, assumed fixed for this calculation. Blue tones denote regions where the planet 
experiences an inward orbital migration, while red tone identify regions where the migration is 
outward. The primordial parameters of Kepler-38 are again highlighted in the top pannel with a 
filled white circle and marked as ``K38''.}
\label{fig7}
\end{figure}

Figure \ref{fig7} shows the normalized value of $\big< da_2/dt \big>$ in the $(\alpha,m_1/m_0)$ 
plane for three different values of the binary and planet eccentricities. For each value of 
$m_1$ the physical radius of the star was modified following the empirical rule ${\cal R_1} 
\simeq 0.9 m_1$. The nominal values are shown in the top panel, and the parameters corresponding 
to Kepler-38 highlighted with a white circle. All initial conditions and physical parameters 
leading to an inward orbital migration of the planet are colored in tones of blue, while those 
leading to a secular increase of $a_2$ in tones of red. The limit between both is marked with a 
white curve.

Although the plots show some quantitative differences as function of the eccentricities, in all 
cases there seems to exist a lower value of $m_1/m_0$ above which the tidal interaction of the 
system leads to an outward migration of the planet. The critical value of $m_1$ appears to be 
larger for more eccentric binaries and lower for stars in almost circular orbits. As expected, 
as $m_1 \rightarrow 0$ the migration is inwards, in accordance with known results for the 
2-body case. 

It is necessary to point out that our analytical model was obtained through a Legendre 
expansion of the elliptic functions truncated at fourth-order of $\alpha$. Consequently, the 
results shown here and in Figure \ref{fig:stat} are not expected to be accurate (or even valid) 
for $\alpha \rightarrow 1$. We have nevertheless opted to include the complete range solely for 
illustrative purposes. 

The time variation of the eccentricity $e_2$ may be found from the orbital angular momentum 
${\bf{L}}_2$ in the Jacobi reference frame. In the planar case, we have
\be
L_2 = \beta_2 |(\boldsymbol{\rho}_{\bf 2} \times \dot{\boldsymbol{\rho}}_{\bf 2})| = \beta_2 
\sqrt{\mathcal{G}\sigma_2 a_2 (1-e^2_2)} ,
\label{eq32}
\ee
whose time derivative due to tidal forces leads to
\be
\frac{1}{\beta_2}\dot{L}_2 = \frac{\mathcal{G}\sigma_2\beta_2}{2 L_2} \bigg( (1-e^2_2) 
\frac{da_2}{dt} - a_2 \frac{de_2^2}{dt} \bigg) = |(\boldsymbol{\rho}_{\bf 2} \times 
\delta{\bf{f}_2})| .
\label{eq33}
\ee
Extracting the eccentricity term, we finally obtain:
\be
\frac{d}{dt} (e^2_2) = \frac{1}{a_2} \bigg[(1-e^2_2)\frac{da_2}{dt} - \frac{2 
L_2}{\mathcal{G}\sigma_2\beta_2} ({\boldsymbol\rho}_{\bf 2} \times \delta\bf{f_2}) \bigg] .
\label{eq34}
\ee
Introducing elliptic expansions in a similar manner as done for (\ref{eq30}), and averaging 
over short-period terms, we obtain:
\be
\bigg< \frac{de_2^2}{dt} \bigg> = \frac{n_2}{4 {\cal G} m_2 \sigma_2 a_2^{5}} 
\sum_{i=0}^4 \sum_{j,k=0}^2 \sum_{l=0}^1 A^{(e)}_{i,j,k,l} \, {\cal K}_{l}^{(+)} 
\gamma_l^i \alpha^i e_1^j e_2^k
\label{eq35}
\ee
where now the non-zero coefficients are given by
\bea
\begin{split}
A^{(e)}_{0,0,2,l} &=  4 \Big[  11 \bar{\Omega}^{(l)}_2 - 18 n_2           \Big] \\
A^{(e)}_{2,0,2,l} &= 20 \Big[  36 \bar{\Omega}^{(l)}_2 + 15 n_1 - 74 n_2  \Big] \\
A^{(e)}_{4,0,2,l} &= 40 \Big[ 139 \bar{\Omega}^{(l)}_2 + 95 n_1 - 282 n_2 \Big] \\
A^{(e)}_{1,1,1,l} &=  2 \Big[  39 \bar{\Omega}^{(l)}_2 - 54 n_2 \Big] \cos(\Delta\varpi) \\
A^{(e)}_{3,1,1,l} &= 10 \Big[ 102 \bar{\Omega}^{(l)}_2 + 34 n_1 - 185 n_2 \Big]  
\cos(\Delta\varpi) .\\
\end{split}
\label{eq36}
\eea
Contrary to $da_2/dt$, we found that the eccentricity of the planet is always damped, at least 
for the initial conditions and system parameters tested here.

\subsection{Comparisons with numerical integrations}

To test the accuracy of our analytical model, for given initial conditions we compare the 
variation in planetary semimajor axis and eccentricity predicted by equations (\ref{eq30}) and 
(\ref{eq35}) with the numerical results obtained using the original unaveraged equations 
(\ref{eq23}) and (\ref{eq34}). We consider the nominal system parameters detailed in Table 
\ref{tab1} but varied the planetary eccentricity and semimajor axis ratio $\alpha$. For each we 
computed $da_2/dt$ and $de_2/dt$ as a function of the reduced mass
\be
\tilde{\mu} = \frac{m_1}{m_0+m_1}
\label{37}
\ee
by varying $m_1$. Due to the rapid rotational synchronization timescales, we consider 
stationary spins for the stars and for the planet according to equation (\ref{eq22}). 
\begin{figure*}
\centering
\includegraphics[width=.95\textwidth,clip=true]{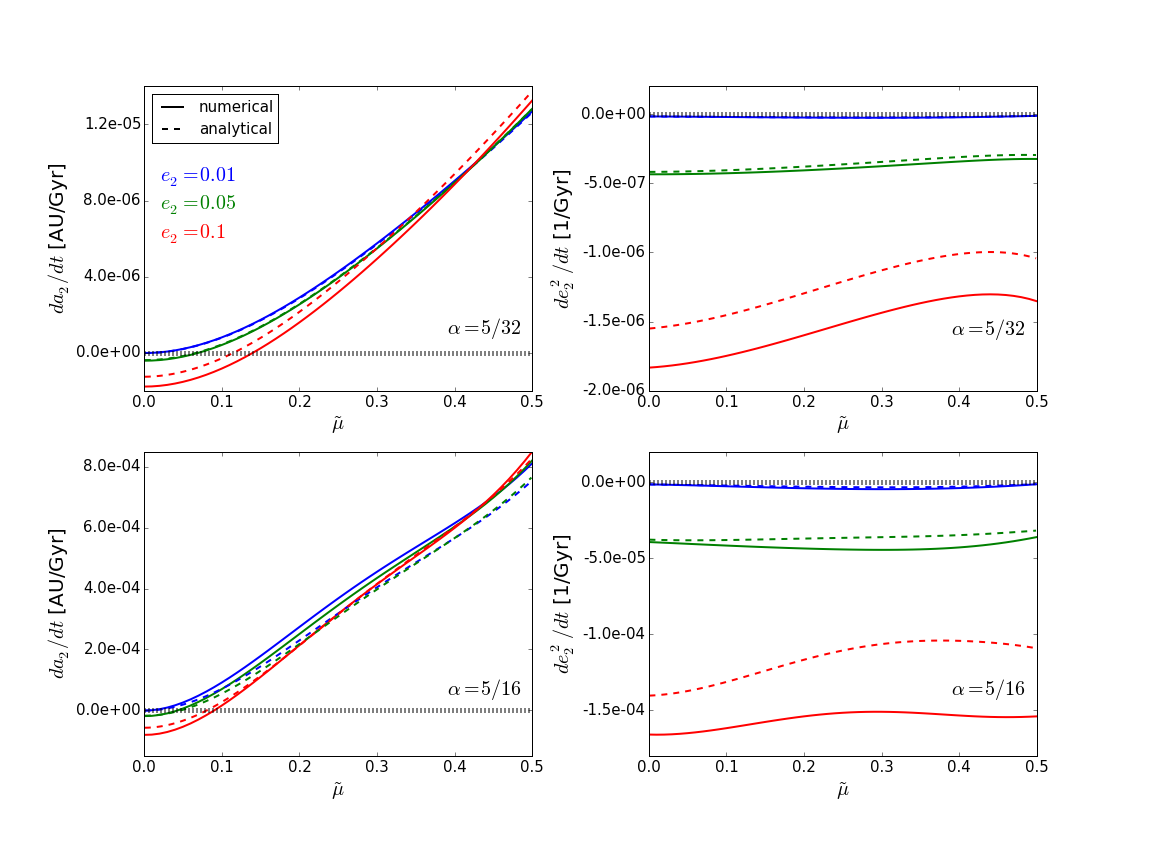}
\caption{Time derivative of the semimajor axis (left panels) and eccentricity variation 
(right panels) of a circumbinary planet at different distances from the binary: $\alpha=5/32$ 
(top panels) and $\alpha=5/16$ (bottom panels). Different colors are employed for different 
eccentricities ($e_2=0.01$ in blue, $e_2=0.05$ in green and $e_2=0.1$ in red) and different type 
of curves make reference to the calculation method: numerical (full line) and analytical (dashed 
line).}
\label{fig:dade}
\end{figure*}

Results are shown in Figure (\ref{fig:dade}). In all the panels the colors represent different 
planetary eccentricities ($e_2=0.01$ in blue, $e_2=0.05$ in green and $e_2=0.1$ in red) while 
the type of curve makes reference to the calculation method (full line for numerical and dashed 
line for analytical). Different rows correspond to different values of $\alpha$: the reference 
value in the bottom panels ($\alpha=5/16$, see Table (\ref{tab1})) and half the nominal value 
in top panels. 

From the right panels we note that, as a result of the tidal interaction, the eccentricity of 
the planet always decreases with a rate that seems weakly dependent on the secondary mass. 
However, as in the 2-body case, $e_2$ decays more rapidly for eccentric planets. Thus, the 
effect of tides on the eccentricity of circumbinary planets is very similar to that in the case 
of bodies around single stars. In the absence of additional forces we expect the systems 
to evolve towards quasi-circular orbits. Since our analytical model only included terms up 
to second order in $e_i$, the accuracy decreases substantially for larger eccentricities, 
leading to an relative error of the order of $20 \%$ for $e_2 \sim 0.1$. A more complete model, 
perhaps including Mignard eccentricity functions (Mignard 1980) are necessary for more 
eccentric orbits. 

The rate of change of the semimajor axis (left-hand plots) shows a better agreement between our 
model and the full unaveraged equations, leading to practically the same magnitude in the 
derivatives even for moderate eccentricities. In particular, the values of the critical reduced 
mass $\tilde{\mu}_{crit}$ associated to the limit between inward and outward migration is very 
well reproduced. 

\begin{figure}
\centering
\includegraphics[width=1.1\columnwidth,clip=true]{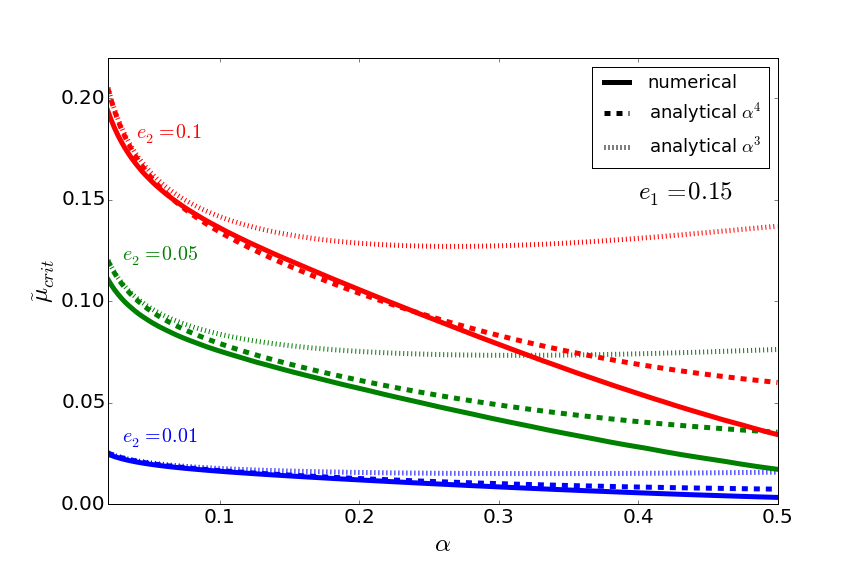}
\caption{Critical value of $\tilde{\mu}$ above which tidal effects on the planet lead to 
outwards orbital migration. Different colors represent different eccentricities for the planet 
(same as in Figure (\ref{fig:dade})) and different types of curves refer to different 
calculation method: numerical (continuous curves), analytical up to fourth order in $\alpha$ 
(dashed) and analytical up to third order in the semimajor-axis ratio (dotted line).}
\label{fig:mucri}
\end{figure}

Finally, Figure (\ref{fig:mucri}) shows the dependence of $\tilde{\mu}_{crit}$ as 
function of $\alpha$ for different eccentricities. As before, calculations performed 
with the unaveraged equations are plotted in continuous lines, while dashed curves 
show results with the analytical model including terms up to fourth order in $\alpha$. To test 
the necessity of such high orders, the dotted lines show analogous results, this time 
truncating the expansions at third order in the semimajor-axis ratio. While the precision of
the fourth-order analytical model is very good up to $\alpha \sim 0.3$, the truncated version 
shows a much smaller region of validity, reduced down to $\alpha \sim 0.1$. Thus, systems such 
as Kepler-38 require a high-order model in order to reproduce the dynamics with a fair accuracy.

It is interesting to note that $\tilde{\mu}_{crit}$ increases for smaller values of $\alpha$. 
In the limit when $\alpha \rightarrow 0$, we expect the system to behave as a planet orbiting a 
single star of mass $m_0+m_1$ and all initial conditions should lead to an inward migration of 
the semimajor axis.

\section{Summary and discussion}
\label{conclu}

In this work we present a model for treating the tides in a circumbinary system with one 
planet, in which all bodies are assumed to be extended and tidally interacting. To built the 
model, we consider a weak friction regime where the tidal forces can be approximated by the 
classical expressions of \cite{Mignard1979} and proceed in two steps:

\begin{enumerate}
\item First, we revisited the Mignard theory and studied which tidal forces have a net effect 
onto the dynamical evolution of the system. In the classical 2-body problem, where we are 
computing the torques on the same body that exerts the deformation, the zero-order Mignard 
torques have zero net secular effect. We found that this torques also has a null effect on 
the third body, as long as there are no mean-motion resonances between $m_1$ and $m_2$. 
Thus, in the non-resonant circumbinary problem, the only forces that should be taken into 
account are those that are applied on the same body that exerts the deformation.
In a resonant case the zero-order torques may have important effects; their consequences will 
be the focus of a forthcoming work. 

\item Secondly, we incorporate the tidal forces in the gravitational equations of motion in a 
self-consistent approach. Namely, we consider that each of the bodies is deformed by the other 
two and there is a reaction force for each tidal force applied. As a result, we obtain the spin 
evolution equation for the bodies and the orbital evolution equation for the planet.
\end{enumerate}

We have undertaken a series of numerical simulations, considering Kepler-38 system as a working 
example, in order to compare the results of this model with our previous work 
(\cite{Zoppetti2018}). We observed that in the short-timescales the dynamics is dominated by the 
spin synchronization of the bodies: the planet, assumed a rocky body, synchronize very quickly 
(in $\sim$ Myr) in a stationary spin lower than the orbital mean motion. On the other hand, the 
stars exhibit super-synchronous spins in values predicted by the 2-body classical problem. The 
subsequent orbital evolution of the binary is little affected by the planet and proceeds to a 
decrease in the semimajor axis $a_1$ and eccentricity $e_1$.

The long-term orbital evolution of the planet is curiously different: as a result of the tidal 
interaction the planet migrates outward and the direction of migration is not dependent on the 
initial planetary eccentricity or the assumed planetary tidal parameter. Moreover, the outward 
migration is also not an indirect effect of the migration of the binary, but observed even if 
the tidal evolution of the stars is neglected. 

During the tidal migration, the eccentricity of the planet oscillates around the force 
eccentricity, which decreases as we move away from the binary (\cite{Leung2013}). For some 
initial conditions, we found that the difference of pericenter angle $\Delta \varpi$ librates 
around zero. Thus, when studying the secular tidal evolution of circumbinary planets, the usual 
procedure of averaging over the longitudes of pericenters may not be accurate.

To better understand the numerical results, we constructed an analytical secular model 
expanding the full spin and orbital equations of motion and averaging only over the mean 
longitudes. Regarding the spins, the simplicity of the full equation, allows us to expand only 
up to second-order in $\alpha$ and the eccentricities $e_1$ and $e_2$. The resulting 
expressions showed a very good agreement with N-body simulations. We furthermore 
obtained a simple equation estimating the stationary spin of CB planets that is not dependent on 
the planetary mass. If we assume that their spins have reached their equilibrium state,
this allow us to predict the rotation period of almost all circumbinary systems requiring only 
knowledge of the stellar masses and the orbital configuration of its members. Our analytical 
approach was validated comparing the planetary stationary spin of the numerical simulation 
with those predicted by our analytical equations.

Contrary to the spins, the analytical model for the orbital evolution required an expansion in 
the semimajor-axis ratio up to fourth-order in $\alpha$. We maintained the eccentricities up to 
second order; however, latter simulations showed that higher orders are probably needed in 
systems with moderate-to-high eccentricities. 

Regarding the eccentricity evolution, we found that the tidal forces on the CB planet always 
seem to act circularizating its orbit. We observed a strong dependence on the eccentricities 
but only a marginal dependence on the mass ratio of the stellar components. On the other hand, 
the complex dependence of the planetary semimajor axis evolution with the mass of the stars is 
reflected in the fact that the direction of migration depends on the binary mass ratio: for 
binaries in which the secondary star is much less massive, even the case in which the secondary 
companion is a planet, the tidal migration direction is inward. However, when the mass of 
both stars are of the same order the planet migrates outward. The critical value of mass ratio 
for which the direction of migration changes sign is dependent on the planetary eccentricity and 
also on the position of the CB planet but can be predicted very accurately with our model.

The magnitude of the semimajor-axis variation is also very sensitive to the planetary 
eccentricity and proximity to the binary, but mainly dominated by the amount of energy that is 
dissipated in the planet due to tides. This quantity is very uncertain; however, the unexpected 
outward tidal migration of CB planet seems to be only dependent on the stellar masses and 
system configuration. A preliminary application of our model to other observed {\it Kepler} 
systems seems to indicate that many systems could also have suffered an outward tidal 
migration.

\begin{acknowledgements}
We wish to express our gratitude to IATE for an extensive use of their computing facilities, 
without which this work would not have been possible. This research was funded by CONICET, 
SECYT/UNC, FONCYT and FAPESP (Grant 2016/20189-9).
\end{acknowledgements}

\bibliographystyle{aa}

\end{document}